\documentclass[12pt,preprint]{aastex}

\shorttitle{NGC 2264 CMM3}
\shortauthors{Watanabe et al.}

\begin{document}


\title{Discovery of Striking Difference of Molecular-Emission-Line Richness in the Potential Proto-Binary System NGC~2264~CMM3}


\author{Yoshimasa~Watanabe\altaffilmark{1,2,3}, Nami~Sakai\altaffilmark{4}, Ana~L\'opez-Sepulcre\altaffilmark{1,5},Takeshi Sakai\altaffilmark{6}, Tomoya~Hirota\altaffilmark{7}, Sheng-Yuan~Liu\altaffilmark{8}, Yu-Nung~Su\altaffilmark{8}}

\and

\author{Satoshi Yamamoto\altaffilmark{1, 9}}
\email{watanabe.yoshimas.ft@u.tsukuba.ac.jp}


\altaffiltext{1}{Department of Physics, The University of Tokyo, 7-3-1 Hongo, Bunkyo-ku, Tokyo, 113-0033, Japan}
\altaffiltext{2}{Division of Physics, Faculty of Pure and Applied Sciences, University of Tsukuba,  Tsukuba, Ibaraki 305-8571, Japan}
\altaffiltext{3}{Center for Integrated Research in Fundamental Science and Engineering (CiRfSE), Faculty of Pure and Applied Sciences, University of Tsukuba,  Tsukuba, Ibaraki 305-8571, Japan}
\altaffiltext{4}{RIKEN, 2-1, Hirosawa, Wako, Saitama 351-0198, Japan}
\altaffiltext{5}{Institut de Radioastronomie Millim\'etrique (IRAM) 38406, Saint Martin d'H\'eres, France}
\altaffiltext{6}{Graduate School of Informatics and Engineering, The University of Electro-Communications, Chofu, Tokyo 182-8585, Japan}
\altaffiltext{7}{National Astronomical Observatory of Japan, Osawa, Mitaka, Tokyo 181-8588, Japan}
\altaffiltext{8}{Academia Sinica, Institute of Astronomy and Astrophysics, PO Box 23-141, Taipei 106, Taiwan}
\altaffiltext{9}{Research Center for the Early Universe, The University of Tokyo, 7-3-1 Hongo, Bunkyo-ku, Tokyo, 113-0033, Japan}


\begin{abstract}
We have conducted an interferometric line survey in the 0.8~mm band toward the young high-mass protostar candidate NGC~2264~CMM3 with ALMA.  CMM3 is resolved into the two continuum peaks, CMM3A and CMM3B, at an angular separation of $0.9''$. Thus, CMM3 is found to be a binary system candidate.  We have detected molecular outflows associated with CMM3A and CMM3B each, indicating active star formation.  In addition to the two peaks, six faint continuum peaks are detected around CMM3A and CMM3B, most of which are thought to be evolved low-mass protostars.  CMM3A is found to be rich in molecular line emission including complex organic molecules such as HCOOCH$_3$ and CH$_3$OCH$_3$.  The emission of complex organic molecules is distributed within a compact region around the continuum peak of CMM3A.  Hence, CMM3A apparently harbors a hot core.  On the other hand, CMM3B is deficient in molecular line emission, although its continuum flux is almost comparable to that of CMM3A.  Possible origins of the striking difference between CMM3A and CMM3B are discussed.
\end{abstract}





\section{Introduction}
NGC~2264~C is a nearby \citep[distance=738~pc][]{Kamezaki2014} cluster forming region.  Many millimeter and submillimeter sources have so far been identified in this region \citep{Ward-Thompson2000,Peretto2006,Peretto2007}.  \citet{Maury2009} found eleven outflow lobes in the NGC~2264~C region by an extensive mapping observation of CO.  Recently, \citet{Ana2016} found the SiO emission with narrow line width in the periphery of NGC~2264~C, and suggested that it reveals a remnant of past star formation activities.  In such a complex structure of this region, NGC~2264~CMM3 is known to be the most massive core \citep[40~M$_{\odot}$:][]{Peretto2006} located almost at its geometrical center.  According to the theoretical model, CMM3 is predicted to evolve into a main sequence star with a mass of $8M_{\odot}$ \citep{Maury2009}.  The protostar of CMM3 is deeply embedded in an envelope, and hence{\rm ,} no mid-infrared source is identified even in the 24~$\mu$m band of \textit{Spitzer}.  A definitive evidence of star formation in CMM3 was first found by \citet{Saruwatari2011}, who identified a compact bipolar outflow with a dynamical age of 140--2000~year by the CO and CH$_3$OH observations with the Submillimeter Array (SMA).  \citet{Cunningham2016} also identified the bipolar outflows in the CO, H$_2$CO, SO, and SiO lines.  Judging from the non-detection of mid-infrared sources and the association of dynamically young outflows, \citet{Saruwatari2011} suggested that CMM3 is a candidate of a high mass protostar in the early evolutionary phase.  

Toward CMM3, a chemical composition has been studied by a spectral line survey observation with single dish telescopes in the 4~mm, 3~mm, and 0.8~mm bands \citep{Watanabe2015}.  In their survey, carbon-chain molecules such as C$_4$H, HC$_5$N, and C$_3$S are found to be relatively abundant in CMM3 compared with the representative high-mass star forming region Orion~KL, while complex organic molecules are deficient.  This result suggests chemical youth of the CMM3 envelope. 
Recently, we carried out a spectral line survey toward CMM3 at a high angular resolution with ALMA in the 0.8~mm band.  Here we report the results on the continuum emission and the distributions of the emission of complex organic molecules.

\section{Observation and Reduction}
The observations were carried out with ALMA on 19 July 2014 and 20 May 2015, where the numbers of the 12~m antenna used in those observations were 31 and 39, respectively.  Minimum and maximum baseline lengths were 14~m and 649~m on the ground, respectively.  Twelve spectral windows were employed to cover the frequency ranges of 336.0--347.3~GHz and 348.0--359.3~GHz at a frequency resolution of 0.98~MHz.  The phase-center coordinate is ($\alpha_{\rm J2000}$, $\delta_{\rm J2000}$) = (6:41:12.3, +9:29:8.0).  The quasars J0510+1800 and J0725-0054 were observed for the bandpass calibration in the 2014 and 2015 sessions, respectively.  The quasars J0643+0857 and J0700+1709 were observed for the amplitude and phase gain calibrations in the 2014 and 2015 sessions, respectively.  The absolute flux was calibrated by using the quasar J0750+125.  The uncertainty of the absolute flux calibration is 10~\% according to ALMA Cycle 2 Technical Handbook (Lundgren~2013).  In these observations, any structure extended over $\sim 13''$ or larger is subject to the resolving-out effect, judging from the minimum baseline length of 14~m.  

Data reduction was carried out with the Common Astronomy Software Applications package (CASA; Version 4.5.3).  A continuum image was obtained by averaging line-free channels of all the spectral windows.  Self-calibration was performed by using the continuum data, and then the solutions were applied to both the continuum and spectral line data.  After the self-calibration procedures, the data were imaged and CLEANed by using the Briggs weighting with the robustness parameter of 0.5.  The resulting synthesized beam and the root-mean-square (rms) noise of the continuum image are $0.37'' \times 0.30''$ (P.A.$=-30.7^{\circ}$) and 0.35~mJy~beam$^{-1}$, respectively.  The signal-to-noise ratio in the continuum image was improved from 190 to 723 by applying the self-calibration procedure.  The synthesized beam and the rms noise of spectral images range from 0.28$''$ to 0.49$''$ and from 4.2~mJy~beam$^{-1}$ to 7.0~mJy~beam$^{-1}$ at the frequency resolution of 0.98~MHz, respectively, depending on the frequency.  

\section{Results}
\subsection{Detection of a Potential Binary System in the CMM3}
\begin{figure}
\epsscale{0.99}
\plotone{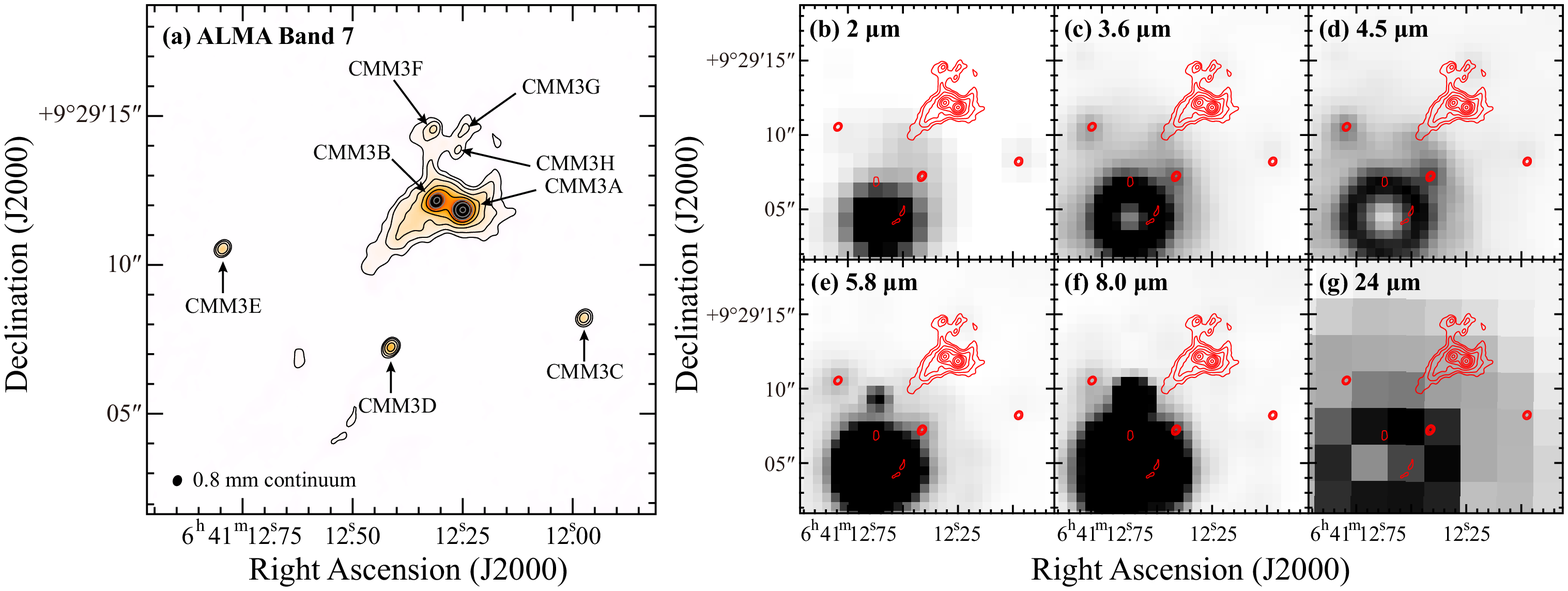}
\caption{(a) The 0.8~mm continuum image observed with ALMA.  Contours levels are $1.7\,{\rm mJy\,beam^{-1}\,}(5 \sigma) \times (1,2,4,8,...,128)$.  CMM3 consists of the two peaks CMM3A and CMM3B.  In addition faint continuum sources CMM3C--CMM3H are found.  Infrared images obtained by (b) 2MASS K band \citep{Skrutskie2006}, (c) $Spitzer$~IRAC 3.6~$\mu$m (PI: John Stauffer), (d) $Spitzer$~IRAC 4.5~$\mu$m (PI: Lucas Cieza), (e) $Spitzer$~IRAC 5.8~$\mu$m (PI: Eric Agol), (f) $Spitzer$~IRAC 8.0~$\mu$m (PI: Giovanni Fazio), and $Spitzer$~MIPS 24~$\mu$m (PI: George Rieke).  The $Spitzer$ data were obtained from NASA/IPAC Infrared Science Archive (http://irsa.ipac.caltech.edu).  Faint infrared emission is seen for CMM3C--CMM3E, while CMM3F--CMM3H is not apparently associated with the infrared source. }
\label{fig01}
\end{figure}

Figure~\ref{fig01}a shows the 0.8~mm continuum image of NGC~2264~CMM3 observed with ALMA.  The beam size of $0.37'' \times 0.30''$ (P.A.$=-30.7^{\circ}$) corresponds to a linear scale of $\sim270$~au at the distance of 738~pc.  CMM~3 is well resolved into the two continuum peaks CMM3A and CMM3B with comparable flux densities of 431.8~mJy and 313.0~mJy, respectively (Table~\ref{tab01}).  The angular separation between the two peaks is $\sim 0.9''$, which corresponds to $\sim660$~au.  The line of sight velocities of molecules associated with CMM3A ($\sim 7.4$~km~s$^{-1}$) are similar to those of CMM3B ($\sim 7.8$~km~s$^{-1}$).  Therefore, the two protostars are most likely gravitationally bound to each other.  In the previous studies, CMM3 has been recognized as a single source because of insufficient angular resolutions \citep[e.g.][]{Peretto2007, Sakai2007, Saruwatari2011,Cunningham2016}.  It should be noted that the potential binary components, CMM3A and CMM3B, are surrounded by an extended envelope. The continuum emission shows asymmetric elongation toward the south-eastern and northern directions from CMM3B and slightly toward the north-western direction from CMM3A.  This emission would likely trace the circumbinary envelope.

In addition to the two bright continuum peaks, six compact continuum sources (CMM3C, CMM3D, CMM3E, CMM3F, CMM3G, and CMM3H) are detected around CMM3A/B, as shown in Figure~\ref{fig01}a.  Here, we have identified local intensity peaks detected with more than $10\sigma$ as the continuum peaks (Table~\ref{tab01}).  CMM3C, CMM3D, and CMM3E are point sources, because the FWHM evaluated by the 2D-Gaussian fitting is similar to the synthesized beam of the observation.  In the near-infrared images (Figures~\ref{fig01}c--f), faint sources seem to be associated with CMM3C, CMM3D, and CMM3E.  Therefore, these three continuum sources would be low-mass Class I or Class II protostars.  On the other hand, CMM3F is marginally resolved, and is associated with no infrared source.  It may represent a low-mass protostellar/prestellar source.  For CMM3G and CMM3H, no parameter was derived by the 2D Gaussinan fitting, because the fitting did not converge.  Note that these two peaks are close to the outflow lobe B1 reported by \citet{Saruwatari2011}.

\subsection{Chemical Difference between CMM3A and CMM3B}
\begin{figure}
\epsscale{0.89}
\plotone{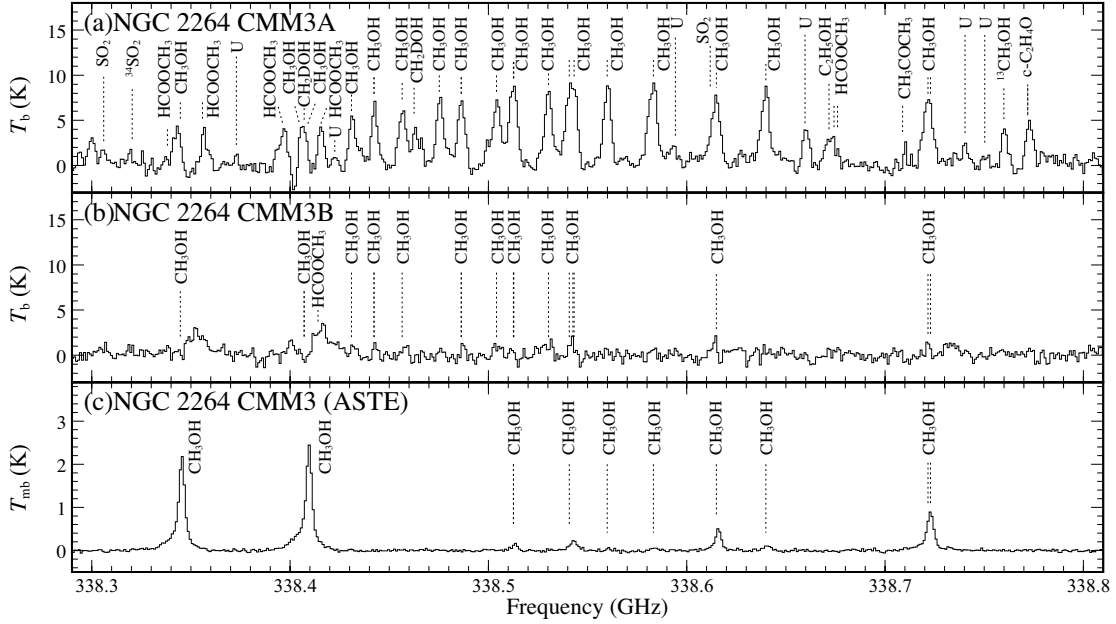}
\caption{\footnotesize (a) The spectrum observed toward CMM3A with ALMA.  (b) The spectrum observed toward CMM3B with ALMA.  Spectral lines are much fainter in CMM3B than in CMM3A.  The intensity scales of ALMA spectra ((a) and (b)) are converted from Jy~beam$^{-1}$ to $T_{\rm b}$ by using the beam size of $0.37'' \times 0.29''$.  The emission lines detected with the confidence level of $3\sigma$ (1.5~K) or higher are labeled.  `U' denotes an unidentified line.  (c) The spectrum observed with the ASTE 10~m telescope.  The single-dish spectrum does not reflect the hot core embedded in CMM3.}
\label{fig02}
\end{figure}

Although CMM3A and CMM3B have comparable flux densities in the 0.8~mm continuum emission, their spectral patterns are found to be much different from each other. Figures~\ref{fig02}a and \ref{fig02}b show the spectra of CMM3A and CMM3B, respectively, in the frequency range from 338.3~GHz to 338.8~GHz.  The spectrum of CMM3A shows many emission lines of CH$_3$OH and SO$_2$, and those of complex organic molecules such as HCOOCH$_3$, CH$_3$CH$_2$OH, and c-C$_2$H$_4$O even in this narrow frequency range.  The full spectra (336.0--347.3~GHz and 348.0--359.3~GHz) and line lists of this spectral line survey will be available in a forthcoming paper (Watanabe~et~al. in prep).  In the previous observation with the single-dish telescope ASTE in the 0.8~mm band (Figure~\ref{fig02}c), the lines of these complex organic molecules were not seen at all except for CH$_3$OH in this frequency range \citep{Watanabe2015}, although HCOOCH$_3$ and CH$_3$OCH$_3$ have been weakly detected in the 3~mm band \citep{Sakai2007,Watanabe2015}.  This means that the hot core was not traced by the  single-dish telescope observation even in the sub-mm band, since its contribution is heavily diluted.  The single-dish spectrum is dominated by the contribution from the surrounding envelope.  For example, only 4~\% of flux density of the CH$_3$OH ($13_1 - 13_0\,{\rm A}^{-+}$) line observed with ASTE is calculated to come from CMM3A and CMM3B.  The high angular resolution observation with ALMA allows us to detect the hot core associated with CMM3A for the first time.  

\begin{figure}
\epsscale{0.89}
\plotone{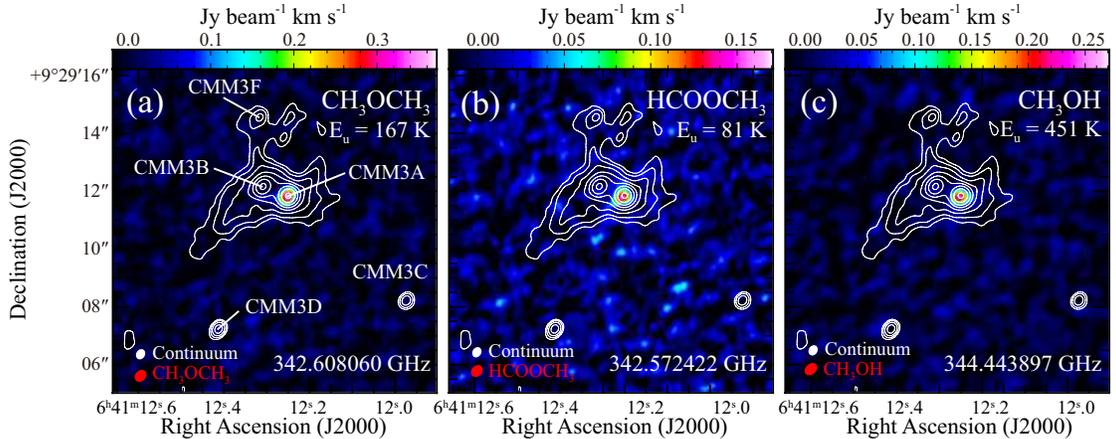}
\caption{\footnotesize  The 0th moment maps of (a) CH$_3$OCH$_3$~($19_{0\,19}-18_{1\,18}\,{\rm AE}$, $19_{0\,19}-18_{1\,18}\,{\rm EA}$, $19_{0\,19}-18_{1\,18}\,{\rm EE}$, and $19_{0\,19}-18_{1\,18}\,{\rm AA}$), (b) HCOOCH$_3$~($11_{8\,4}-10_{7\,3}\,{\rm A}$ and $11_{8\,3}-10_{7\,4},{\rm A}$), and CH$_3$OH~($19_1-18_2\,{\rm A^+}$) superposed on the contours of the continuum image.  The velocity range of the integration is from 3~km~s$^{-1}$ to 10~km~s$^{-1}$ for these transition lines.  The contours levels of the continuum image are the same as those in Figure~\ref{fig01}.  The distributions of these molecules are concentrated in CMM3A. }
\label{fig03}
\end{figure}

\begin{figure}
\epsscale{0.50}
\plotone{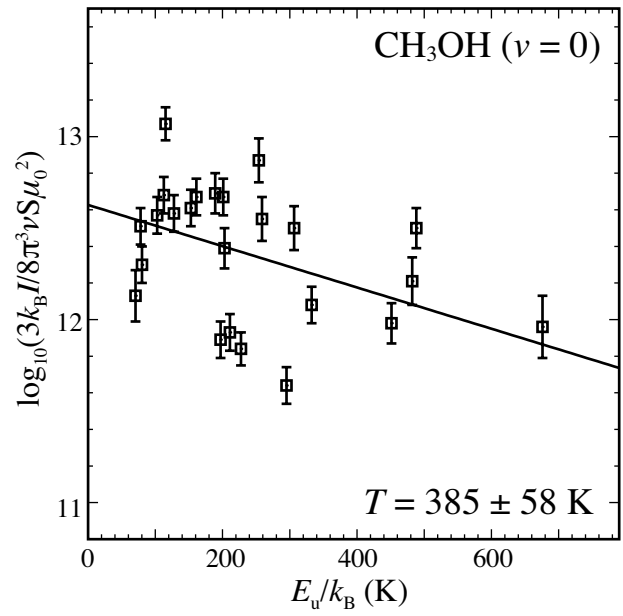}
\caption{\footnotesize  The rotation diagram plots for CH$_3$OH.  The vibrationally excited transition lines and the lines blended with other transition lines are excluded from the plots.  The rotation temperature is derived to be $385 \pm 58$~K. }
\label{fig04}
\end{figure}

Figures~\ref{fig03}a, b and c show integrated intensity maps of CH$_3$OCH$_3$, HCOOCH$_3$, and CH$_3$OH, respectively.  These transition lines are selected as examples to demonstrate the distributions of complex organic molecules and that of CH$_3$OH with the very high upper-state energy (451~K), because they seem to be less contaminated from other molecular lines.  As expected from their spectra (Figure~\ref{fig02}a,b), the complex organic molecules and the very high excitation CH$_3$OH are associated only with CMM3A.  Furthermore, they are concentrated in a compact region ($<300$~au) around the continuum peak.  High excitation lines ($E_{\rm u}>100$~K) of complex organic molecules are detected, and hence, a hot core is thought to have already been formed in CMM3A.  The existence of the hot core is indeed supported by the high rotation temperature ($385 \pm 58$~K) of CH$_3$OH estimated by using the rotation diagram method under the assumption of optically thin conditions and local thermodynamic equilibrium (LTE) (Figure~\ref{fig04}).  The uncertainty of the derived rotation temperature is relatively large because of the non-LTE effects, optical depths of lines, blending with the other lines, and the resolving-out effect for low-excitation lines.  Nevertheless, the high temperature condition is evident.  

In contrast to the richness of emission lines in CMM2A, only CH$_3$OH is detected marginally in CMM3B.  Although the HCOOCH$_3$ may marginally be detected, it could be blended with the CH$_3$OH line.  Since no complex organic molecules are definitively detected (Figure~\ref{fig03}a, b), a hot core has not yet been formed in CMM3B.  

We here evaluate column densities of CH$_3$OH, CH$_3$OCH$_3$, and HCOOCH$_3$, and their fractional abundances relative to H$_2$.  The column densities are derived under the LTE approximation with the rotation temperatures of 100~K and 300~K.  Because CH$_3$OCH$_3$ and HCOOCH$_3$ are not definitively detected in CMM3B, their column densities cannot be determined for CMM3B.  In order to compare the chemical compositions between CMM3A and CMM3B on the same basis, we derived the column densities by using the same transitions.  As mentioned above, the rotation temperature evaluated from the rotation diagram of CH$_3$OH (Figure~\ref{fig04}) suffers from various systematic uncertainties.  Therefore, we employ the rotation temperatures of 100~K and 300~K for both CMM3A and CMM3B.  We use a few transitions without contamination from other molecular lines as listed in the captions of Table~\ref{tab02}, and evaluate the column densities by taking the averages of the results from the transition lines.  The results are summarized in Table~\ref{tab02}.  

The column densities of H$_2$ ($N_{\rm H_2}$) are calculated from the continuum peak flux of the 0.8~mm band by using the formula used by \citet{Imai2016} with the dust temperatures of 100~K and 300~K.  Here, we adopt the mass absorption coefficient of the dust of $\kappa_{\nu}=0.0182$~cm$^2$g$^{-1}$ derived by \citet{Jorgensen2016}, which is an interpolated value of the numerical simulation \citep{Ossenkopf1994}.  The gas-to-dust mass ratio is assumed to be 100.  

The derived H$_2$ column densities and fractional abundances of molecule relative to H$_2$ are summarized in Table~\ref{tab02}.  The fractional abundances of CH$_3$OH and CH$_3$OCH$_3$ are found to be higher in CMM3A than those in CMM3B by factor of 5, while the upper limit of HCOOCH$_3$ is not stringent.  This chemical difference will be discussed in the latter section. 

The HCOOCH$_3$/CH$_3$OH and CH$_3$OCH$_3$/CH$_3$OH ratios in CMM3A are roughly comparable to or even higher than the averaged values in the low-mass and high-mass protostars \citep{Taquet2015}.  Therefore, CMM3A possesses analogous chemical compositions of complex organic molecules to other protostars.

\subsection{Outflows}
\begin{figure}
\epsscale{0.99}
\plotone{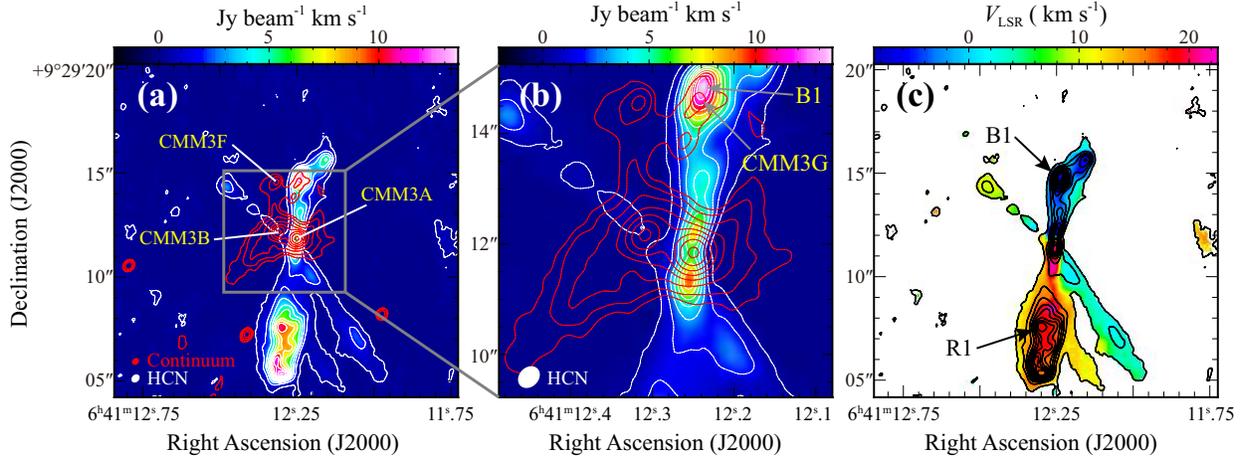}
\caption{(a) The 0th moment maps of HCN and (b) its blow-up, where the velocity range of the integration is from -28~km~s$^{-1}$ to 42~km~s$^{-1}$.  The red contours represent the continuum distribution.  The contours levels of the continuum image are the same as those in Figure~\ref{fig01}.  The white contours represent the integrated intensity of HCN.  Contours levels are $0.13\,{\rm Jy\,beam^{-1}\,km\,s^{-1}\,}(1 \sigma) \times (3,13,23,...,93)$.   The two outflow systems can be seen; one along the almost north to south direction and another along the northeast to southwest direction.  (c) The 1st moment map of the two outflows.  The 1st moment map is calculated by using the pixels with $> 3\sigma$ (15~mJy~beam$^{-1}$) in the same velocity range of the 0th moment.  Hence, the region of which 0th moment is lower than $3\sigma$ are clipped.  The black contours represent the integrated intensity of HCN.  The contours levels are the same as those in Figure~\ref{fig01}a.  B1 and R1 indicate the positions of CO lobes identified by \citet{Saruwatari2011}.}
\label{fig05}
\end{figure}

\begin{figure}
\epsscale{0.99}
\plotone{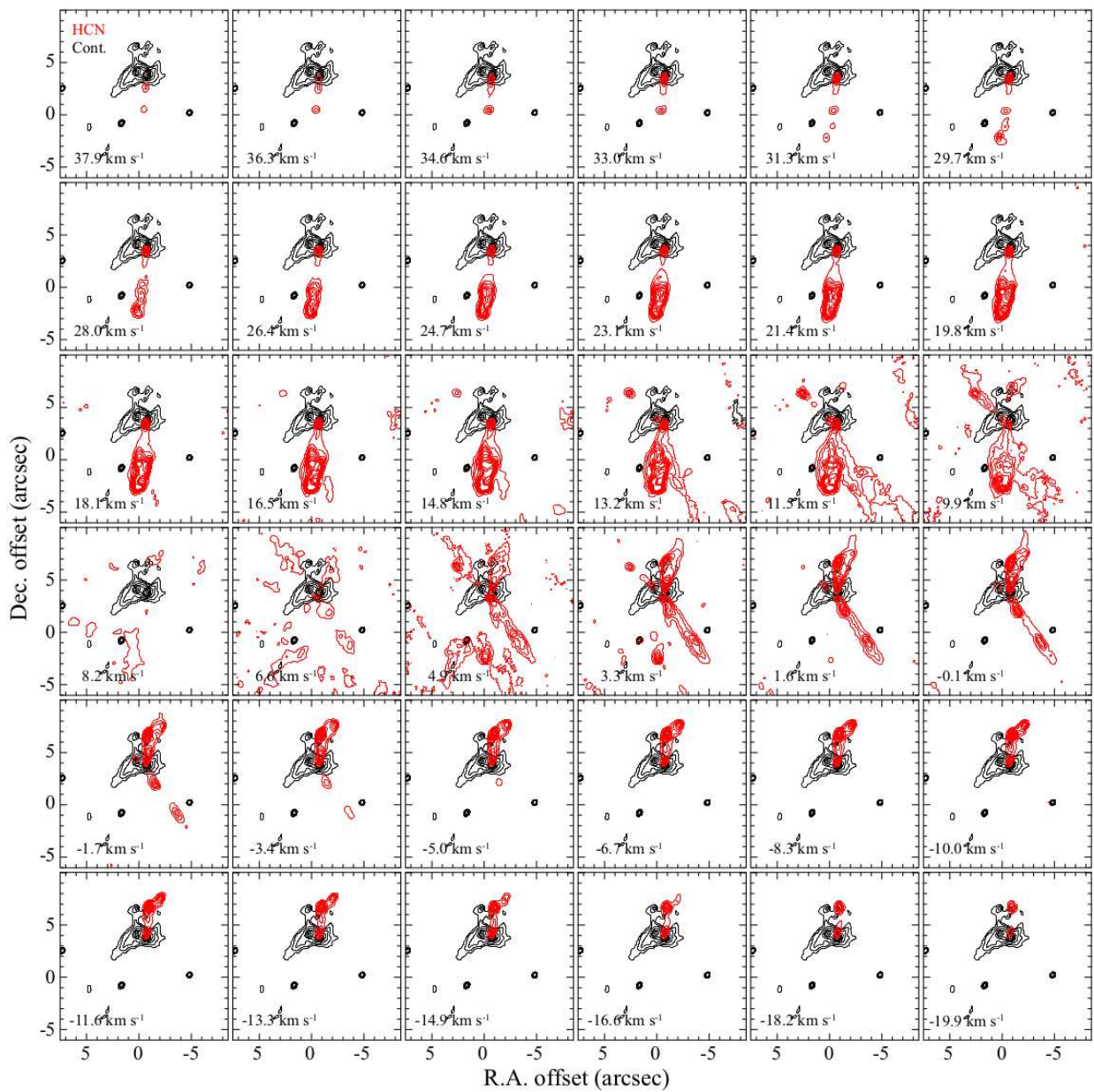}
\caption{Velociity channel maps of HCN (red contours).  The contour levels represent $5\,{\rm mJy\,beam^{-1}}\,(1\sigma) \times (5, 15, 25, ..., 95)$.  The black contours are the continuum distribution.  The contour levels are the same as those in Figure~\ref{fig01}.  The coordinates represent the offset from the phase center ($\alpha_{\rm J2000}$, $\delta_{\rm J2000}$) = (6$^{\rm h}$41$^{\rm m}$12$^{\rm s}$.3, +9$^{\circ}$29$'$8$''$.0).}
\label{fig06}
\end{figure}

\begin{figure}
\epsscale{0.99}
\plotone{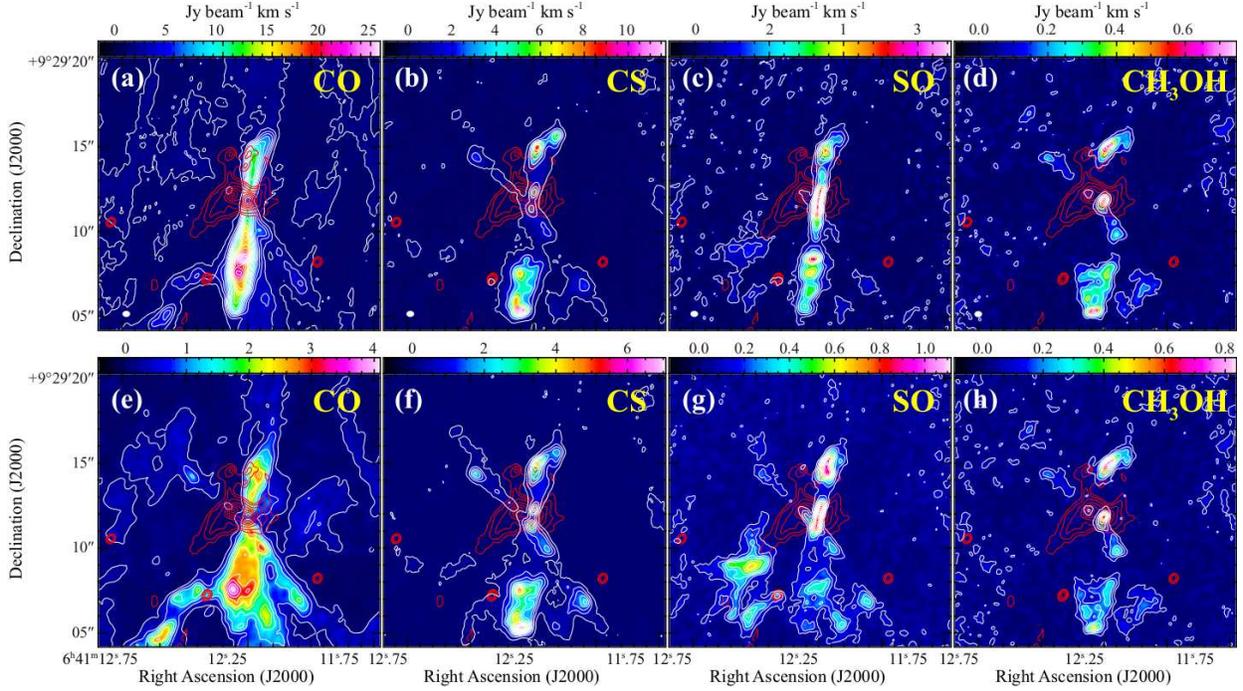}
\caption{The 0th moment maps of (a) CO (3--2), (b) CS (7--6), (c) SO($J_N = 7_8-6_7$), and (d) CH$_3$OH ($13_1-13_0\,{\rm A}^{-+}$).  The velocity ranges for the integration are from -28~km~s$^{-1}$ to 43~km~s$^{-1}$,  from -26~km~s$^{-1}$ to 39~km~s$^{-1}$, from -25~km~s$^{-1}$ to 41~km~s$^{-1}$, and from -15~km~s$^{-1}$ to 36~km~s$^{-1}$ for CO, CS, SO, and CH$_3$OH, respectively.  The red contours represent the continuum distribution.  The contour levels of the continuum image are the same as those in Figure~\ref{fig01}.  The white contours represent the integrated intensity of the molecules.  The contour levels are $0.13\,{\rm Jy\,beam^{-1}\,km\,s^{-1}\,}(1 \sigma) \times (3,13,23,...,193)$, $0.12\,{\rm Jy\,beam^{-1}\,km\,s^{-1}\,}(1 \sigma) \times (3,13,23,...,93)$, $0.04\,{\rm Jy\,beam^{-1}\,km\,s^{-1}\,}(1 \sigma) \times (3,8,13,...,53)$, and $0.02\,{\rm Jy\,beam^{-1}\,km\,s^{-1}\,}(1 \sigma) \times (3,8,13,...,33)$ for CO, CS, SO, and CH$_3$OH, respectively.  The 0th moment maps of (e) CO (3--2), (f) CS (7--6), (g) SO($J_N = 7_8-6_7$), and (h) CH$_3$OH ($13_1-13_0\,{\rm A}^{-+}$) derived with a narrower velocity range of the integration than those of (a), (b), (c), and (d), respectively.  The velocity range for the integration is from -4~km~s$^{-1}$ to 17~km~s$^{-1}$ for these figures in order to reveal the second outflow.  The red contours represent the continuum distribution with the same contours levels in Figure~\ref{fig01}.  The white contours represent the integrated intensity of molecules.  The contours levels are $0.07\,{\rm Jy\,beam^{-1}\,km\,s^{-1}\,}(1 \sigma) \times (3,9,12,...,57)$, $0.06\,{\rm Jy\,beam^{-1}\,km\,s^{-1}\,}(1 \sigma) \times (3,11,19,...,115)$, $0.03\,{\rm Jy\,beam^{-1}\,km\,s^{-1}\,}(1 \sigma) \times (3,6,9,...,36)$, and $0.02\,{\rm Jy\,beam^{-1}\,km\,s^{-1}\,}(1 \sigma) \times (3,7,11,...,36)$ for CO, CS, SO, and CH$_3$OH, respectively.  }
\label{fig07}
\end{figure}

Figures~\ref{fig05}a and c show the 0th and the 1st moment maps of HCN, respectively.  We find a prominent bipolar outflow blowing from CMM3A toward northern and southern directions, which are blue-shifted and red-shifted, respectively (Figure~\ref{fig05}c).  The blue-shifted and red-shifted outflows can also be seen in the channel maps of HCN (Figure~\ref{fig06}).  This bipolar outflow system corresponds to that found by \citet{Saruwatari2011} in the CO($J=2-1$) line with SMA.  Their B1 and R1 lobes are seen in Figure~\ref{fig05}, while B2 and R2 lobes are out of the field of view in the present ALMA observation.  As seen in the expanded figure of the 0th HCN moment map (Figure~\ref{fig05}b), this bipolar outflows system is launched from CMM3A.  It is also visible in other molecular lines including CO~($J=3-2$), CS~($J=7-6$), SO~($N_J=8_7-7_6$), and CH$_3$OH~(e.g., $13_1-13_0, {\rm A^{-+}}$) (Figure~\ref{fig07}a-d).   

It should be noted that the continuum peak CMM3G is close to the B1 lobe (Figure~\ref{fig02}).  This indicates that the blue-shifted outflow is interacting with a dense material traced by the continuum emission (CMM3G) and the B1 lobe is induced by the shocks caused by the interaction.  Moreover, we find that the outflow is elongated toward the northwest direction from the B1 lobe (Figure~\ref{fig05}a).  This elongation can be seen from 3.3~km~s$^{-1}$ to -16.6~km~s$^{-1}$ in the channel maps (Figure~\ref{fig06}).  Therefore, the direction of outflow may be changed by the dynamical interaction with the dense clump of CMM3G. 

In addition to the prominent bipolar outflow, we find another bipolar outflow system along the northeast to southwest direction (Figures~\ref{fig05}a).  The second outflow is also seen from 16.5~km~s$^{-1}$ to -5.0~km~s$^{-1}$ in the channel maps of HCN (Figure~\ref{fig06}).  This outflow system is well-collimated, and is also visible in other molecular lines including CO, CS, SO, and CH$_3$OH (Figure~\ref{fig07} e-h).  The velocity-integrated emission from the second outflow is much fainter than that from the prominent outflow, because the velocity range of the second outflow is narrower ($\sim 20$~km~s$^{-1}$ than that of the prominent outflow ($\sim 70$~km~s$^{-1}$).  Moreover, a part of the second outflow is strongly affected by the resolving-out effect at the system velocity (8.2~km~s$^{-1}$ and 6.6~km~s$^{-1}$ in Figure~\ref{fig06}).  The southwestern lobe of this outflow can be seen in the CH$_3$OH ($5_0 - 4_0$ A$^{+}$) line in Figure~4a of \citet{Saruwatari2011}.  This collimated outflow system seems to be launched from CMM3B, since the outflow axis just passes through the peak of CMM3B (Figure~\ref{fig05}b).  The velocity field map (Figure~\ref{fig05}c) shows  small velocity gradient along the bipolar outflow axis.  This result indicates that the outflow blows almost in parallel to the plane of sky.  The detection of this outflow provides us with a strong evidence of the presence of a protostar in CMM3B.  Its well collimated feature suggests that the outflow system would most likely be in the early phase \citep{Arce2006}.

\section{Discussion}
In this study, we have found that CMM3 is a potential binary protostellar source.  More importantly, the binary components, CMM3A and CMM3B, show different spectral appearance in spite of their comparable continuum intensities.  It is interesting to note that a similar feature has recently been reported for the young low-mass binary source NGC~1333~IRAS4A \citep{Ana2017}.  We here discuss possible origins of the difference in CMM3. 

The first possibility is the difference of the physical evolutionary stage between the two sources.  Considering the prominent outflow feature and the association of the hot core, CMM3A seems to harbor young high-mass (or intermediate-mass) protostar.  In contrast, deficient molecular emission in CMM3B might originate from its youth.  Namely, the protostar may be so young that the hot core would not have been developed yet or would be still tiny around the protostar.  In this case, most of organic molecules are still frozen out on dust grains, and their gas phase abundances are expected to be low.  This possibility is supported by the well collimated outflow from CMM3B, which is usually interpreted as a sign of a young dynamical age \citep{Arce2006}.  The envelope mass is derived based on the formula given by \citet{Ward-Thompson2000}.  For CMM3A, it is evaluated to be 18~M$_{\odot}$ and 5.7~M$_{\odot}$, assuming the dust temperature of 100~K and 300~K, respectively.  For CMM3B, it is 13~M$_{\odot}$ and 4.1~M$_{\odot}$, respectively.  Thus, the envelope mass of CMM3B could be higher than CMM3A, if the dust temperature is lower in CMM3B because of its youth.

The second possibility is the dust opacity effect.  The dust emission at 0.8~mm is not always optically thin at a scale of a few tens of au.  If the dust opacity is so high that the emission of various molecules originated from the innermost part of the dust emitting region is heavily absorbed by the dust, the molecular emission could be very weak.  This may be the situation for CMM3B.  Since the dust temperature of CMM3A is expected to be higher than that of CMM3B, the dust opacity would be lower in CMM3A, and this effect is less significant.  As mentioned before, it is most likely that the outflow from CMM3B is almost on the plane of sky.  In this case, the disk/envelope system of CMM3B is almost edge-on, which could make the dust opacity high.  

The third possibility is the difference of masses of the protostars between CMM3A and CMM3B.  If CMM3B is less massive than CMM3A, a size of the hot region where complex organic molecules can be sublimated is very small due to low luminosity.  Since the hot region is smaller in CMM3B than CMM3A, the dust temperature is also expected to be lower in CMM3B than CMM3A.  Therefore, a larger envelope gas would still be remaining in CMM3B, as discussed in the first possibility.  

Apparently, these three possibilities are interrelated to one another.  The scenario of a less massive protostar for CMM3B is related to the evolutionary scenario, because the protostar evolution would strongly depend on its mass.  It is also related to the dust opacity scenario, because a young protostar is deeply embedded in the parent cloud.  In order to disentangle these three possibilities, we need to define the spectral energy distributions for the two sources.  Then, we can derive the disk/envelope mass accurately, and estimate the evolutionary stage of the protostar reasonably.  High angular resolution observations resolving CMM3A and CMM3B at various wavelength from cm wave to sub-mm wave are awaited.






\acknowledgments
We thank the anonymous reviewer for helpful comments and suggestions to improve this paper.  This paper makes use of the ALMA data set ADS/JAO.ALMA\#2013.1.01192.S.  ALMA is a partnership of the ESO (representing its member states), the NSF (USA) and NINS (Japan), together with the NRC (Canada) and the NSC and ASIAA (Taiwan), in cooperation with the Republic of Chile. The Joint ALMA Observatory is operated by the ESO, the AUI/NRAO and the NAOJ. The authors are grateful to the ALMA staff for their excellent support.  This publication makes use of data products from the Two Micron All Sky Survey, which is a joint project of the University of Massachusetts and the Infrared Processing and Analysis Center, funded by the National Aeronautics and Space Administration and the National Science Foundation.  This research has made use of the NASA/ IPAC Infrared Science Archive, which is operated by the Jet Propulsion Laboratory, California Institute of Technology, under contract with the National Aeronautics and Space Administration.  This study is supported by a Grant-in-Aid from the Ministry of Education, Culture, Sports, Science, and Technology of Japan (No. 25108005 and 16K17657).



{\it Facilities:} \facility{ALMA}.


\clearpage


\clearpage
\begin{table}
\caption{Continuum Peak Parameters of NGC2264~CMM3 \label{tab01}}
\scriptsize  
\begin{tabular}{cllllll}
\tableline\tableline
            & R.A. (J2000) & Dec. (J2000) & Flux \tablenotemark{a} & Peak Flux & Size \tablenotemark{b} & Mass \tablenotemark{c} \\
            &  &  &  (mJy)  & (mJy~beam$^{-1}$) & (arcsec) & (M$_{\odot}$)\\
\tableline
CMM3A &  6$^{\rm h}$41$^{\rm m}$12.25$^{\rm s}$ & +9$^{\circ}$29$'$11.85$''$ & $431.8 (2.2)$ & $253.2$ & $0.34 \times 0.27$ (PA$=68^{\circ}$)  & 18 (2) - 5.7 (0.6) \\
CMM3B &  6$^{\rm h}$41$^{\rm m}$12.31$^{\rm s}$ & +9$^{\circ}$29$'$12.16$''$ & $313.0 (3.4)$ & $128.9$ & $0.57 \times 0.45$ (PA$=125^{\circ}$) & 13 (1) - 4.1 (0.4)\\
CMM3C &  6$^{\rm h}$41$^{\rm m}$11.97$^{\rm s}$ & +9$^{\circ}$29$'$8.22$''$  &  $12.7 (0.5)$ & $ 11.7$ & \tablenotemark{d} & 0.53 (0.06) - 0.17 (0.02)\\
CMM3D &  6$^{\rm h}$41$^{\rm m}$12.41$^{\rm s}$ & +9$^{\circ}$29$'$7.23$''$  &  $21.7 (0.8)$ & $ 21.0$ & \tablenotemark{d} & 0.9 (0.1) - 0.29 (0.03)\\
CMM3E &  6$^{\rm h}$41$^{\rm m}$12.80$^{\rm s}$ & +9$^{\circ}$29$'$10.55$''$ &  $13.6 (0.7)$ & $ 13.5$ & \tablenotemark{d} & 0.57 (0.06) - 0.18 (0.02)\\
CMM3F &  6$^{\rm h}$41$^{\rm m}$12.32$^{\rm s}$ & +9$^{\circ}$29$'$14.40$''$ &  $36.5 (6.4)$ & $  8.6$ & $0.93 \times 0.69$ (PA$=21^{\circ}$) & 1.5 (0.3) - 0.5 (0.1) \\
CMM3G \tablenotemark{e} & 6$^{\rm h}$41$^{\rm m}$12.24$^{\rm s}$ & +9$^{\circ}$29$'$14.60$''$ & \tablenotemark{f}   & $  4.1$ & &  \\
CMM3H \tablenotemark{e} & 6$^{\rm h}$41$^{\rm m}$12.32$^{\rm s}$ & +9$^{\circ}$29$'$14.40$''$ & \tablenotemark{f}   & $  4.0$ & &  \\
\tableline
\end{tabular}
\tablenotetext{a}{Flux density by 2D Gaussian fit.}
\tablenotetext{b}{Deconvolved source size by 2D Gaussian fit.}
\tablenotetext{c}{Envelope mass of CMM3A, CMM3B, CMM3C, CMM3D, CMM3E, and CMM3F.  The envelope mass are evaluated from the flux density derived by the 2D Gaussian fit at the dust temperatures of 100~K and 300~K by using the formula used in \citet{Imai2016}.  Here, the mass absorption coefficient of dust and the gas-to-dust ratio are assumed to be  $0.0182$~cm$^2$g$^{-1}$ \citep{Jorgensen2016} and 100, respectively.} 
\tablenotetext{d}{Deconvolved source size cannot be calculated.  It is a point source.}
\tablenotetext{e}{The coordinates are derived from the pixel of the local intensity peak because the 2D Gaussian fit did not converge.}
\tablenotetext{f}{The flux density cannot be derived, because the 2D Gaussian fit did not converge.}
\end{table}

\begin{table}
\caption{Column Densities of Molecules at CMM3A and CMM3B \label{tab02}}
\scriptsize  
\begin{tabular}{llllll}
\tableline\tableline
              & & \multicolumn{2}{c}{CMM3A} & \multicolumn{2}{c}{CMM3B} \\
Molecule      & $T$ (K) \tablenotemark{a} & $N$ (cm$^{-1}$) \tablenotemark{bc} & $X$ \tablenotemark{cd}      & $N$ (cm$^{-1}$) \tablenotemark{bc} & X \tablenotemark{cd} \\
\tableline
H$_2$                           & 100 & $5.1 (0.5) \times 10^{25}$             &               
                                      & $2.6 (0.3) \times 10^{25}$             &      \\
                                & 300 & $1.6 (0.2) \times 10^{25}$             &               
                                      & $8.1 (0.8) \times 10^{24}$             &      \\
CH$_3$OH \tablenotemark{e}      & 100 & $4.0 (0.2) \times 10^{16}$             & $7.8 (0.9) \times 10^{-10}$ 
                                      & $4.0 (0.8) \times 10^{15}$             & $1.6 (0.4) \times 10^{-10}$     \\
                                & 300 & $4.6 (0.3) \times 10^{16}$             & $2.9 (0.3) \times 10^{-9}$ 
                                      & $4.5 (0.9) \times 10^{15}$             & $6 (1) \times 10^{-10}$     \\
CH$_3$OCH$_3$ \tablenotemark{f} & 100 & $6.0 (0.5) \times 10^{15}$             & $1.2 (0.2) \times 10^{-10}$ 
                                      & $< 4 \times 10^{14}$ \tablenotemark{g} & $< 2 \times 10^{-11}$ \tablenotemark{g} \\
                                & 100 & $7.6 (0.6) \times 10^{15}$             & $4.8 (0.6) \times 10^{-10}$ 
                                      & $< 7 \times 10^{14}$ \tablenotemark{g} & $< 8 \times 10^{-11}$ \tablenotemark{g} \\
HCOOCH$_3$ \tablenotemark{h}    & 100 & $2.5 (0.2) \times 10^{16}$             & $4.9 (0.5) \times 10^{-10}$ 
                                      & $< 4 \times 10^{15}$ \tablenotemark{g} & $< 2 \times 10^{-10}$ \tablenotemark{g} \\
                                & 300 & $6.7 (0.5) \times 10^{16}$             & $4.2 (0.4) \times 10^{-9}$ 
                                      & $< 1 \times 10^{16}$ \tablenotemark{g} & $< 2 \times 10^{-9}$ \tablenotemark{g}  \\
\tableline
\end{tabular}
\tablenotetext{a}{Assumed temperature.}
\tablenotetext{b}{Column density of molecule.  The column density of H$_2$ is evaluated by assuming dust temperatures of 100~K and 300~K.  The column density of other molecules are evaluated by assuming rotation temperatures of 100~K and 300~K.}
\tablenotetext{c}{Errors are evaluated by taking into account the rms noise ($1\sigma$) and calibration uncertainty (10\%).}
\tablenotetext{d}{Fractional abundance relative to the H$_2$ molecule.}
\tablenotetext{e}{The lines used are $7_{6}-6_{6}$ A$^+$ (338.442344~GHz) and $7_{6} - 6_{6}$ A$^+$ (338.442344~GHz), and $7_{-5}-6_{-5}$ E (338.456499~GHz) and $7_{5}-6_{5}$ E (338.475290~GHz) for the A and E species, respectively.  The column densities of the A and E species are derived by averaging the results for the A and E species lines, respectively.  Then, the total column densities of CH$_3$OH are evaluated by summing the column densities of A and E species.} 
\tablenotetext{f}{The lines used are $19_{0\,19}-18_{1\,18}$ AA (342.608044~GHz), $19_{0\,19}-18_{1\,18}$ EE (342.607971~GHz), $19_{0\,19}-18_{1\,18}$ AE (342.607898~GHz) , $19_{0\,19}-18_{1\,18}$ EA (342.607898~GHz), $18_{2\,17}-17_{1\,16}$ AA (357.460920~GHz), $18_{2\,17}-17_{1\,16}$ EE (357.460164~GHz), $18_{2\,17}-17_{1\,16}$ EA (357.459408~GHz), and $18_{2\,17}-17_{1\,16}$ AE (357.459408~GHz).}
\tablenotetext{g}{Upper limit is evaluated from the $3\sigma$ upper limit of the integrated intensity assuming a line width of 2~km~s$^{-1}$.}
\tablenotetext{h}{The lines used are $11_{8\,4}-10_{7\,3}$ A (342.572422~GHz), $12_{8\,5}-11_{7\,4}$ A (354.805708~GHz), and $12_{8\,4}-11_{7\,5}$ A (354.805708~GHz), and $12_{8\,5}-11_{7\,5}$ E (354.742476~GHz) and $17_{6\,12}-16_{5\,11}$ E (355.312450~GHz) for the A and E species, respectively.  The column densities of the A and E species are derived separately by averaging the results for the A and E species lines, respectively.  Then, the total column densities of HCOOCH$_3$ are evaluated by summing the column densities of the A and E species.} 
\end{table}

\end{document}